# Effective Refractive-Index Approximation: A Link between Structural and Optical Disorder of Planar Resonant Optical Structures

Žarko Gačević[1,*] and Nenad Vukmirović[2]

[1]*ISOM, Universidad Politécnica de Madrid, Avenida Complutense s/n, 28040 Madrid, Spain*
[2]*Scientific Computing Laboratory, Center for the Study of Complex Systems, Institute of Physics Belgrade, University of Belgrade, Pregrevica 118, 11080 Belgrade, Serbia*

We provide detailed insights into a link between structural and optical disorder of resonant optical structures, in particular, distributed Bragg reflectors (DBRs) and resonant microcavities ($\mu$Cs). The standard (targeted) DBR structures have periodic square-wave–like refractive-index profiles, and their optical performance is determined by the refractive-index ratio of the two applied materials ($n_{12} = n_1/n_2$, $n_1 > n_2$) and the number of DBR periods ($N$). It is well known that its structural disorder strongly affects its optical properties, but, despite that, this influence has not been quantitatively addressed in the literature. We propose a precise quantitative definition for a structural disorder of a single DBR unit cell (disorder factor $D_F$), completing the set of DBR fundamental parameters ($n_{12}$, $N$, $D_F$). Then we expose the basis for the effective refractive-index approximation (ERIA), showing that, as long as DBR optical properties are concerned, the influence of increasing structural disorder ($D_F\uparrow$) is virtually identical to the influence of decreasing refractive-index ratio ($n_{12}\downarrow$), with the latter influence being easily quantified. Making use of the ERIA method, simple analytical formulas, which enable rapid insights into the reflectivity and stop-band width of DBRs with different types of transient layers at the heterointerfaces, are derived and the results validated, via both transfer-matrix simulations and direct experimental measurements of imperfect DBRs. The insights of the ERIA method are then further applied on resonant $\mu$Cs, providing a comprehensive link between their structural disorder and subsequent deterioration of their quality ($Q$) factor.

DOI: 10.1103/PhysRevApplied.9.064041

## I. INTRODUCTION

Distributed Bragg reflectors (DBRs) are fundamental building blocks of numerous optoelectronic devices, such as resonant-cavity light-emitting diodes [1], vertical-cavity surface-emitting lasers [2], polariton lasers [3], and semi-conductor saturable absorber mirrors [4]. The requirements to achieve DBR high peak reflectivity and a wide stop band combine (i) a high refractive-index ratio between the two employed materials $n_{12} = n_1/n_2 (n_1 > n_2)$, (ii) a high number of DBR periods $N$, and (iii) high structural quality, that is, flat and abrupt DBR interfaces [5]. The combination of materials with different structural and mechanical properties often constrains the formation of flat and abrupt interfaces resulting in the final structure deviation from the targeted one. This issue (commonly referred to as structural disorder) causes undesired photon scattering (commonly referred to as optical or photonic disorder) leading to deterioration of DBR optical performance.

While the influence of (i) refractive-index ratio and (ii) number of periods on DBR optical performance is clear (and easy to calculate), the influence of (iii) structural disorder remains largely unaddressed in the literature. One of the reasons for the last item is that there is no consensus on a clear quantitative definition for DBR structural disorder, thus impeding quantitative insights.

The origins of structural disorder of DBRs can be roughly divided into two main groups: (i) random thickness fluctuations, affecting interface flatness, and (ii) formation of transient layers (TLs) at the DBR heterointerfaces (intermixing), affecting interface abruptness [6]. To get insights into optical properties of DBRs with or without TLs at the interfaces, several methods have been reported in the literature, such as transfer-matrix simulations (TMSs) [5], coupled-mode theory [7,8], and the hyperbolic tangent substitution technique [9]. Concerning the DBRs with TLs at the interfaces, TMSs represent the most frequently applied method. This method, via discretization of the refractive index in the graded transient regions, determines DBR optical properties (i.e., its peak reflectivity and stop-band width). This approach, however, does not provide comprehensive insights into the link between the DBR structural disorder and its optical deterioration or a deeper qualitative understanding of the calculated stop-band parameters.

[*]Corresponding author.
gacevic@isom.upm.es

In this work, we propose a precise quantitative definition of the DBR unit-cell structural disorder (disorder factor $D_F$). With this definition, we complete the set of quantitative parameters which fully determine the DBR optical performance: $n_{12}$, $N$, and $D_F$. Given a DBR with a certain degree of structural disorder ($n_{12}$, $N$, $D_F > 0$), we further show that the effect of increase in structural disorder ($D_F\uparrow$) on DBR optical properties is practically identical to the effect of a decrease in refractive-index ratio ($n_{12}\downarrow$), with the latter influence being easily quantified. This observation is the essence of the effective refractive-index approximation (ERIA) method and allows any imperfect DBR ($n_{12} > 1$, $N$, $D_F > 0$) with a nearly constant structural disorder ($D_F = $ const) to be substituted with its standard-DBR counterpart (DBR') with a perfect structure but a reduced refractive-index ratio (i.e., $n'_{12} < n_{12}$, $N' = N$, $D'_F = 0$). Consequently, the ERIA method allows rapid calculations of optical properties and a deeper qualitative understanding of disordered DBRs. The results obtained by the method are validated via both TMSs and direct experimental measurements of imperfect DBRs.

DBRs are widely used in different resonant optical devices, perhaps most noticeably in resonant microcavities ($\mu$Cs). Over the last half century, huge advances have been made in the field of solid-state lighting, thanks mainly to the development of a wide variety of direct-band-gap III-V and II-VI materials. The potential of the technology has been demonstrated by the commercialization of light-emitting diodes and edge-emitting laser diodes in a wide wavelength range (from ultraviolet to infrared) [10]. In recent decades, significant efforts have been dedicated to extend the field of edge-emitting devices to the field of vertical-cavity ones. However, two important problems arise when device design changes from an edge-emitting to a vertical-cavity one.

The first problem is related to injection of electrical current through usually thick DBRs which surround the device active region. The conduction- and valence-band profiles of standard (perfect) DBRs are often characterized by high band offsets at the interfaces, which can fully block the current flow through the structure. This problem is particularly important in the case of vertical-cavity surface-emitting laser structures, and, to surmount it, several groups have reported that, with intentional grading of DBRs, i.e., the intentional introduction of TLs at their interfaces, the electrical resistance of the vertical structure can be significantly reduced [11–14]. This improvement, however, is at the cost of reduced DBR reflectivity and, consequently, a reduced quality ($Q$) factor of the resonant $\mu$C.

The second problem is related to $Q$-factor precise control. Note that the $Q$ factor directly affects properties such as the spectral purity of resonant-cavity light-emitting diodes [15], the threshold current of vertical-cavity surface-emitting lasers [16], and the exciton-photon coupling of polariton lasers [17]. The influence of cavity disorder on dispersion in cavity resonant wavelength, cavity detuning, and polariton broadening, as well as the cavity $Q$ factor, has been studied in different materials from III-V and II-VI groups, with all studies confirming that cavity disorder has a strong impact on each of these parameters [18–24]. In the specific case of the $Q$-factor studies, several authors reported that, with sufficiently small excitation spots (a few microns in size), the impact of structural disorder on measurement can be avoided, leading to experimental $Q$ values in close agreement with the theoretical ones [19,21]. Despite that (and similar to the case of DBRs), the link between the two remains unexplained.

In this article, the insights of the ERIA method are further exploited to provide a comprehensive link between the resonant $\mu$C structural disorder and the deterioration of its $Q$ factor. Obtained results are validated via TMSs of resonant $\mu$Cs containing DBRs with graded interfaces.

## II. OPTICAL PROPERTIES OF DBRs

Before a detailed analysis of imperfect DBRs is given, a brief explanation concerning a standard (perfect) DBR design is given for clarity.

### A. A standard DBR: Perfect synchronization of reflected components

A standard DBR structure is designed in a way which maximizes its reflectivity at a given (targeted) wavelength. When monochromatic light with the targeted wavelength hits the DBR structure perpendicularly, all reflected components (created at each DBR interface) reflect back "in phase." The perfect synchronization of the reflected components is the driving mechanism for the maximization of the DBR reflectivity at this specific wavelength [see an example of a GaN/AlN DBR structure in Fig. 1(a)], and it relies on the following three rules:

 (i) Each layer has a quarter-wavelength optical thickness. The wave passage forward and backward through a single DBR layer thus brings the total phase gain of $\pi$.
 (iii) According to the Fresnel law, the refraction at every odd interface ($n_{\text{GaN}} > n_{\text{AlN}}$) brings no phase gain, whereas refraction at every even interface ($n_{\text{AlN}} < n_{\text{GaN}}$) brings $\pi$ phase gain.
 (iii) According to the Fresnel law, there is no phase gain due to wave transmission at the DBR interfaces.

The combination of (i)–(iii) allows for perfect synchronization of all reflected components (it can be easily shown that the components reflected after multiple reflections also satisfy the perfect synchronization condition).

The main optical properties of a DBR are summarized in three stop-band parameters: stop-band position (targeted wavelength, $\lambda$), stop-band height (peak reflectivity, $R$), and stop-band width ($\Delta\lambda/\lambda$). In the case of a standard DBR, these three parameters can be calculated making use of the following formulas:

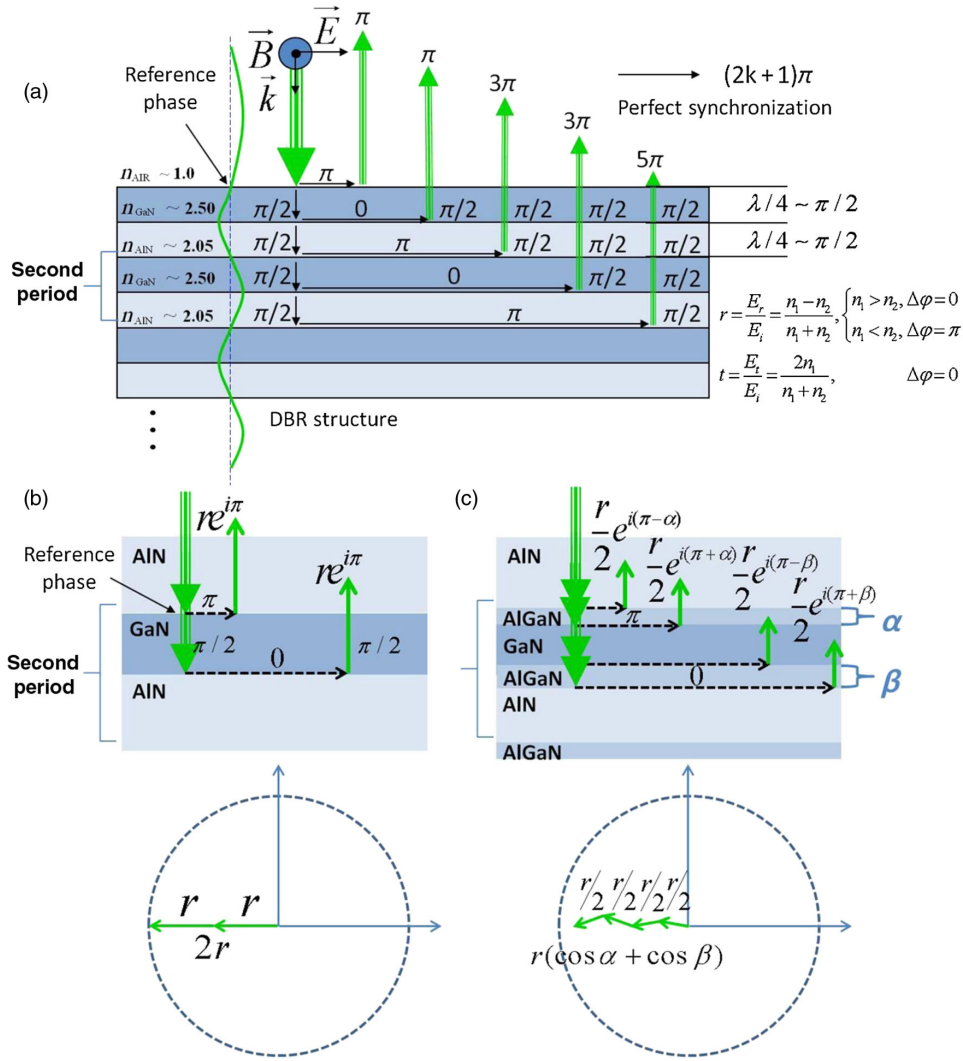

FIG. 1. (a) Perfect synchronization of reflected components of an incident monochromatic light with the targeted wavelength $\lambda$. Perfect synchronization is achieved via DBR design, taking care of the phase gain due to wave propagation, reflection, and transmission. Fresnel coefficients for perpendicular reflection ($r$) and transmission ($t$) are shown, for clarity. (b) Components reflected across one unit cell of an ideal GaN/AlN DBR and their corresponding phasors. (c) Components reflected across one unit cell of a nonideal GaN/AlN DBR with homogeneous transient layers at the interfaces and their corresponding phasors. Phase thicknesses of the transient layers are designated, for clarity. (b) and (c) demonstrate that the origin of decrease in peak reflectivity in the latter structure is desynchronization of the reflected components [25].

$$\lambda = 4n_1 d_1 = 4n_2 d_2, \quad (1)$$

$$R = \left(\frac{1 - \frac{n_s}{n_0} n_{12}^{2N}}{1 + \frac{n_s}{n_0} n_{12}^{2N}}\right)^2, \quad n_{12} = n_1/n_2, \quad (2)$$

$$\frac{\Delta\lambda}{\lambda} = \frac{4}{\pi} \sin^{-1}\left|\frac{n_{12} - 1}{n_{12} + 1}\right|, \quad n_{12} = n_1/n_2, \quad (3)$$

where $n_1$ and $n_2$ ($d_1$ and $d_2$) are the first and second quarter-wave layer refractive indices (thicknesses), respectively (corresponding to the GaN and AlN layers in Fig. 1, respectively), $n_0$ and $n_s$ are the surrounding media refractive indices (commonly air and substrate, respectively), $n_{12}$ is the refractive-index ratio ($n_1/n_2$), and $N$ is the number of DBR periods (stacks) [5]. Note that the stop-band height (peak reflectivity) of a DBR placed between two media with known refractive indices ($n_0$ and $n_s$) is a two-parameter function (a function of refractive-index ratio $n_{12}$ and the number of periods $N$). Concerning the formula for the stop-band width, it, strictly speaking, holds for DBRs with an infinite number of periods ($N = \infty$); however, it is commonly used to calculate the stop-band width of DBRs with an arbitrary number of periods since it is a very good approximation. Therefore, the stop-band width is a one-parameter function, depending solely on the refractive-index ratio $n_{12}$.

### B. An imperfect DBR: Desynchronization of reflected components

Real DBRs deviate (slightly or significantly) from their targeted design. This deviation is commonly referred to as the structural disorder. Note that Eqs. (2) and (3) express the peak reflectivity and stop-band width of perfect DBRs. The equations take into consideration only the refractive-index ratio ($n_{12}$) and the number of DBR periods ($N$), but they do not account for DBR structural disorder, which may have a significant impact on its optical performance.

As previously commented, the main origin of structural disorder is the deterioration of DBR interfaces, in terms of either flatness or abruptness [6]. The interface deterioration leads to a "weakening" of the resonant condition resulting in the creation of reflected components which are no longer perfectly synchronized. This desynchronization affects the DBR optical performance, reducing its peak reflectivity and stop-band width; the stop-band position ($\lambda$), on the other hand, remains (nearly) unaffected [25].

To account for the impact of structural disorder on DBR peak reflectivity and stop-band width, a clear quantitative definition of it is necessary first.

### 1. Structural disorder of a DBR unit cell: A quantitative approach

Let us suppose that, due to technological limitations the fabrication of a DBR, which is designed as shown in Fig. 1(b) (standard DBR) results in a structure with homogeneous TLs at the interfaces, as shown in Fig. 1(c) (imperfect DBR).

Next, let us assume (for simplicity) that the refractive-index variation ($\Delta n = n_{max} - n_{min}$) along the DBR is relatively small, i.e., $[(n_{max} - n_{min})/(n_{max} + n_{min})] \ll 1$. Bearing this limitation in mind, we approximate the average refractive index $\bar{n}$ of a DBR simply as $\bar{n} = [(n_{max} + n_{min})/2]$. A small variation of refractive index along the DBR structure further implies that

 (i) the intensity of the wave vector $k$ is nearly constant along the DBR structure, $k = [(2\pi\bar{n})/\lambda]$.

Next, according to the Fresnel law, the reflection coefficient ($r$) at the interface is $r = [(n_1 - n_2)/(n_1 + n_2)]$ and is much smaller than 1 (for example, for the GaN/AlN interface, this value is $r \approx 0.1$). This result has two important consequences. In a first-approximation analysis (see Fig. 1 for clarity),

 (ii) the contribution of multiple reflections to the total reflectivity can be neglected (since $r \gg r^2 \gg r^3 \gg \ldots$), and

 (iii) the incident-wave intensity is nearly constant along one DBR unit cell [the reflected intensity is proportional to $r^2$ ($r^2 \approx 0.01$), implying that approximately 99% of the incident power is transmitted across one GaN/AlN interface].

According to the adopted approximations (i)–(iii), in the case of the targeted (perfect) DBR, the wave reflected across one unit cell consists of two major (directly reflected) components [see the sketch and the phasor diagram in Fig. 1(b)]. These two reflected components have the same intensity ($r$) and phase ($\pi$), and they yield total reflection coefficient $r_0$, which we refer to as the "unit-cell optical response": $r_0 = -2r$ [Fig. 1(b); the negative sign is due to the phase].

In the case of the resulting (imperfect) DBR with homogeneous TLs [with refractive indices $n_{TL} = [(n_1 + n_2)/2]$ and "phase thicknesses" of $\alpha$ and $\beta$, respectively; see Fig. 1(c)], the reflected wave consists of four (major) directly reflected components. The four reflected components have nearly the same intensity [approximately equal to $r/2$; see the Fresnel coefficients in Fig. 1(a) and the depiction in Fig. 1(c)] but different phases, yielding a total reflection coefficient, i.e., a unit-cell optical response of $r_1 = -r(\cos\alpha + \cos\beta)$ [see Fig. 1(c)] [25]. As can be seen when comparing the two phasor diagrams [Figs. 1(b) and 1(c)], the resulting reflected wave in the latter case has the same phase ($\pi$) but an attenuated intensity. Bearing in mind that the latter (imperfect) structure is an undesired result when trying to fabricate the former (perfect) structure, this undesired attenuation can be used as a quantitative measure of the DBR unit-cell structural ideality, or simply as its ideality factor ($I_F$):

$$I_F = \frac{r_1}{r_0} = \frac{1}{2}(\cos\alpha + \cos\beta). \quad (4)$$

Similarly, for a quantitative measure of the structural disorder of the resulting DBR unit cell, i.e., its disorder factor ($D_F$), the following definition can be used:

$$D_F = \frac{r_0 - r_1}{r_0 + r_1}. \quad (5)$$

It is easy to show that

$$D_F = \frac{1 - I_F}{1 + I_F} \quad (6)$$

and that an ideal unit cell [that of the perfect DBR, Fig. 1(b)], has the ideality factor 1 ($I_F = 1$) and disorder factor zero ($D_F = 0$). In general, $1 \geq I_F > 0$, while $0 \leq D_F < 1$.

### 2. Effective refractive-index approximation: A DBR with homogeneous transient layers (discrete case)

Note that the formation of TLs affects only the intensity (not the phase) of the reflected wave [Fig. 1(c)]; in that sense, the same optical response would be obtained on a unit cell with a reduced refractive-index ratio $n'_{12}$ ($n'_{12} < n_{12}$). To get insight into the optical properties of the resulting imperfect DBR (with structural disorder), we can, therefore, approximate the resulting structure to a standard DBR (zero structural disorder) with the same number of periods ($N$) but a reduced refractive-index ratio $n'_{12}$ by keeping the optical responses of the two DBR unit cells identical; in other words, the response of one unit cell of the DBR with homogeneous TLs ($r_1$) should match that of its standard DBR counterpart (DBR', $r'_0$): $r_1 = r'_0$. It is easy to show that this reduced refractive-index ratio $n'_{12}$, which we will refer to as the effective refractive-index ratio of the DBR under study, is

$$n'_{12} = \frac{(1+n_{12}) - I_F(1-n_{12})}{(1+n_{12}) + I_F(1-n_{12})} = n_{12} \frac{1 + D_F/n_{12}}{1 + D_F n_{12}}. \quad (7)$$

Since the unit-cell optical thickness should be preserved, the following relation holds, $n'_1 d_1 + n'_2 d_2 = n_1 d_1 + n_2 d_2$, which, combined with Eqs. (1) and (7), yields refractive indices of the equivalent standard-DBR counterpart:

$$n'_1 = n_1 \frac{2}{1 + \frac{1 + D_F \frac{n_1}{n_2}}{1 + D_F \frac{n_2}{n_1}}}, \qquad n'_2 = n_2 \frac{2}{1 + \frac{1 + D_F \frac{n_2}{n_1}}{1 + D_F \frac{n_1}{n_2}}}, \quad (8)$$

expressed as a function of the refractive indices and disorder factor of the DBR under study [25]. Note that, by applying the ERIA method, the original four-layer DBR, with parameters $n_1$, $n_2$, and $D_F > 0$, is substituted for its standard-DBR counterpart, with parameters $n'_1$, $n'_2$, and $D'_F = 0$. The two structures have practically identical optical properties (as confirmed in Sec. III). Thus, the relevant optical properties of the imperfect DBR under study can be directly calculated by making use of its standard-DBR counterpart, making use first of Eq. (7) (to find the effective refractive-index ratio $n'_{12}$) and then of Eqs. (2) and (3) (describing perfect DBRs):

$$R = \left( \frac{1 - \frac{n_s}{n_0} n_{12}^{2N} \left( \frac{1 + D_F/n_{12}}{1 + D_F n_{12}} \right)^{2N}}{1 - \frac{n_s}{n_0} n_{12}^{2N} \left( \frac{1 + D_F/n_{12}}{1 + D_F n_{12}} \right)^{2N}} \right)^2, \quad n_{12} = n_1/n_2, \quad (9)$$

$$\frac{\Delta \lambda}{\lambda} = \frac{4}{\pi} \sin^{-1} \left| \frac{1 - D_F}{1 + D_F} \frac{n_{12} - 1}{n_{12} + 1} \right|$$
$$= \frac{4}{\pi} \sin^{-1} \left| I_F \frac{n_{12} - 1}{n_{12} + 1} \right|, \qquad n_{12} = n_1/n_2. \quad (10)$$

Equations (9) and (10) calculate the peak reflectivity and stop-band width of any imperfect DBR with a (nearly) constant structural disorder, making use of the complete set of its fundamental parameters ($n_{12}$, $N$, $D_F$).

### 3. Effective refractive-index approximation: A DBR with graded transient layers (continuous case)

The previous results can be generalized to DBRs with arbitrarily graded interfaces (i.e., DBRs containing TLs with a continuously changing refractive index) by discretizing the refractive index down to the monolayer scale. Assuming that $M$ is the total number of monolayers contained in one DBR unit cell, that $n(m)$ is the refractive index of the $m$th monolayer, and that $I_0$ is the intensity of the incident wave (which remains nearly constant along one DBR period), we can calculate the intensity of the wave reflected across one DBR unit cell $I_{r1}$:

$$I_{r1} = \sum_{m=0}^{M-1} \frac{n(m-1) - n(m)}{n(m-1) + n(m)} e^{2i(m/M)\pi} I_0. \quad (11a)$$

Bearing the adopted approximations in mind, considering the refractive index as a function of the incident-wave phase $n = n(\varphi)$, the sum of discrete components can be written in a more convenient integral form leading to the DBR unit-cell optical response

$$r_1 = \frac{I_{r1}}{I_0} = \int_0^\pi \frac{-n'(\varphi) d\varphi}{2\bar{n}} e^{2i\varphi}, \quad (11b)$$

where $n'(\varphi)$ denotes the first derivative of the function $n(\varphi)$.

Recalling, further, that a unit cell of a standard DBR, with refractive-index ratio $n_1/n_2$ has the optical response $r_0 = (I_{r0}/I_0) = -2r \approx -[(n_1 - n_2)/\bar{n}]$, we arrive at the unit-cell optical response attenuation, i.e., the ideality factor of the resulting structure in the case of graded TLs:

$$I_F = \frac{r_1}{r_0} = \frac{1}{2(n_1 - n_2)} \int_0^\pi n'(\varphi) e^{2i\varphi} d\varphi. \quad (12)$$

Once the ideality and disorder factors are known, the set of relevant DBR parameters is closed ($n_{12}$, $N$, and $D_F$), allowing calculation of its peak reflectivity and stop-band width via Eqs. (9) and (10), respectively.

A DBR unit-cell ideality factor is a function of the DBR TL properties, i.e., their thicknesses and their refractive-index profiles. Similar to the quarter-wave-layer thickness ($\lambda/4$), which can be expressed in terms of the targeted wavelength phase change upon propagation ($\pi/2$), both TL thicknesses ($\alpha$ and $\beta$) and their refractive-index profile [$n(\varphi)$] can be represented as a function of phase (which is convenient for mathematical analysis). Table I depicts and represents analytically refractive-index profiles of DBRs with no TLs [standard (ST) DBR], and with homogeneous ($H$), linearly graded ($L$), sine-wave graded (SW), biparabolically graded (BP), uniparabolically graded (UP), and interdiffused (ID) TLs [9,11–14,26]. The presence of TLs ($H$, $L$, SW, BP, UP, and ID) attenuates the optical response of a single DBR unit cell, with respect to that of the ST one. The attenuation, i.e., the resulting unit-cell ideality factor is calculated in each of the mentioned cases as a function of TL thicknesses ($\alpha$ and $\beta$) and the specific refractive-index profile [$n(\varphi)$] via Eq. (12) (see Appendix A for details). Finally, for quantitative comparison, numerical examples for specific TL thicknesses $\alpha = \beta = 60° = \pi/3$ are also given, for clarity.

## III. VALIDATION OF THE ERIA METHOD

### A. Theoretical validation

For theoretical validation of the ERIA method, a 20-period GaN/AlN DBR, with refractive-index ratios of 2.50/2.05, centered at 405 nm, is considered. The targeted structure thus has 20 stacks of 40.5 nm and 49.4 nm GaN/AlN layers. We further assume that, due to interface intermixing, different types of TLs are formed at the interfaces (we note that the TLs can also be intentionally grown), resulting in (i) homogeneous, (ii) linear, (iii) sine-wave, and (iv) interdiffused profiles, as presented in Table I. For simulations, the phase thicknesses of the transient layers are set to $\alpha = \beta = 60° = \pi/3$ for structures (i)–(iii). For structure (iv) the diffusion length is set to $L_D = 6$ nm. The targeted refractive-index profile and the resulting one, for each of the structures (i)–(iv), are shown on the left-hand side of Figs. 2(a)–2(d).

TABLE I. DBRs with different TL types. No TLs: standard (ST) DBR. Homogeneous (H) TLs, linearly (L), sine-wave (SW), biparabolically (BP), and uniparabolically (UP) graded TLs, as well as interdiffused (ID) TLs. The TLs are both depicted and shown in their analytical form, for clarity. Ideality factors, as a function of TL properties (thicknesses and refractive-index profiles) are calculated analytically. Numerical values of ideality and disorder factors and effective refractive-index ratios (for specified TL thicknesses) are stated for comparison.

| TL types depicted | TL profile: $n(\varphi)$ First TL analytically presented | $I_F(\alpha,\beta)$ $\alpha = \beta = 60° = \pi/3$ | GaN/AlN DBR $I_F$ | GaN/AlN DBR $D_F$ | Effective refractive-index ratios |
|---|---|---|---|---|---|
| ST | ... | $\frac{1}{2}(1+1) = 1$ | 1.000 | 0.000 | 2.500/2.050 |
| H | $n(\varphi)^{+(\alpha/2)}_{-(\alpha/2)} = [(n_1+n_2)/2]$ | $\frac{1}{2}(\cos\alpha + \cos\beta)$ | 0.500 | 0.333 | 2.371/2.156 |
| L | $n(\varphi)^{+(\alpha/2)}_{-(\alpha/2)} = [(n_1+n_2)/2]$ $+ (n_1-n_2)(\varphi/\alpha)$ | $\frac{1}{2}\{[(\sin\alpha)/\alpha] + [(\sin\beta)/\beta]\}$ | 0.827 | 0.095 | 2.457/2.085 |
| SW | $n(\varphi)^{+(\alpha/2)}_{-(\alpha/2)} = [(n_1+n_2)/2]$ $+ [(n_1-n_2)/2]\sin(\pi/\alpha)\varphi$ | $\frac{1}{2}\{[[\cos\alpha/(1-4\alpha^2/\pi^2)]$ $+[\cos\beta/(1-4\beta^2/\pi^2)]\}$ | 0.900 | 0.053 | 2.475/2.071 |
| BP | $n(\varphi)^{\alpha}_0 = n_1 - (n_1-n_2)[(\alpha-\varphi)/\alpha]^2$ | $\frac{1}{2}\{[[\sqrt{\alpha^2 - \alpha\sin(2\alpha)} + (\sin\alpha)^2]/\alpha^2\}$ $+\{[\sqrt{\beta^2 - \beta\sin(2\beta)} + (\sin\beta)^2]/\beta^2\}$ | 0.884 | 0.062 | 2.471/2.074 |
| UP | $n(\varphi)^{\alpha}_0 = n_1 - (n_1-n_2)[(\alpha-\varphi)/\alpha]^2$ | $I_F(\alpha)$ $\alpha = 60° = \pi/3$ $(\beta = 0° = 0)$ $\frac{1}{2}\{[\sqrt{\alpha^2 - \alpha\sin(2\alpha)} + (\sin\alpha)^2/\alpha^2] + 1\}$ | 0.942 | 0.030 | 2.486/2.062 |
| ID | See Appendix A [26] | $I_F(L_d)$ $[L_d/(\lambda/4\bar{n})] = 0.1$ $e^{-(4\pi\bar{n}L_d/\lambda)^2}$ | 0.906 | 0.049 | 2.477/2.069 |

The reflectivity of each of the structures (i)–(iv) is determined by the TMSs [5]. Then the properties of their equivalent standard DBRs are found by making use of the ERIA method, and the reflectivity profile of these DBRs is calculated by the TMS as well. The two reflectivity profiles are compared on the right-hand side of each panel in Figs. 2(a)–2(d). As can be seen, the two reflectivity profiles are in virtually perfect agreement, thus validating that the ERIA method is convenient for an estimation of both peak reflectivity and stop-band width of imperfect (but highly periodic) DBR structures.

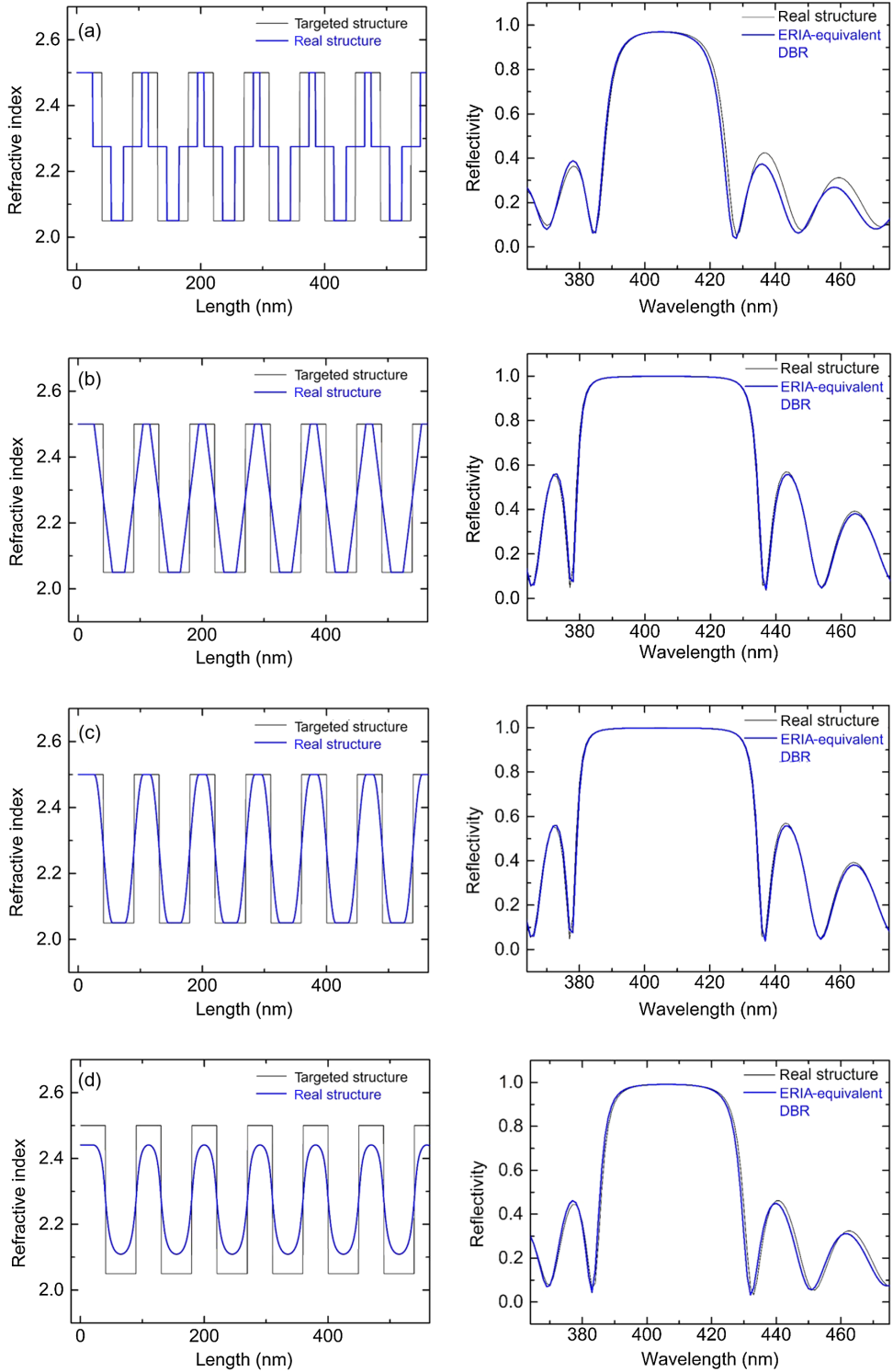

FIG. 2. Theoretical validation of the ERIA method. A 20-period GaN/AlN DBR, with refractive-index ratio of 2.50 / 2.05 centered at 405 nm, is considered a targeted structure. The resulting (real) structures have transient layers at the interfaces in the (a) homogeneous, (b) linear, (c) sine-wave, or (d) interdiffused form (left panels; only the first six periods are shown, for clarity). The right panels compare reflectivity profiles of the real structures with those obtained on their standard-DBR counterparts determined via the ERIA method.

### B. Experimental validation

Figure 3(a) represents the results obtained by measuring the reflectivity of a 20-period GaN/AlN DBR, centered at 405 nm, grown by plasma-assisted molecular beam epitaxy in a Riber Compact 21 reactor. The targeted structure has 20 stacks of 40.5 nm and 49.4 nm GaN/AlN layers (for growth details, see Gačević et al. [25]). The reflectivity spectra of obtained DBRs are measured at room temperature at nearly normal incidence (approximately equal to 8°) with a commercial JASCO V-630 spectrophotometer.

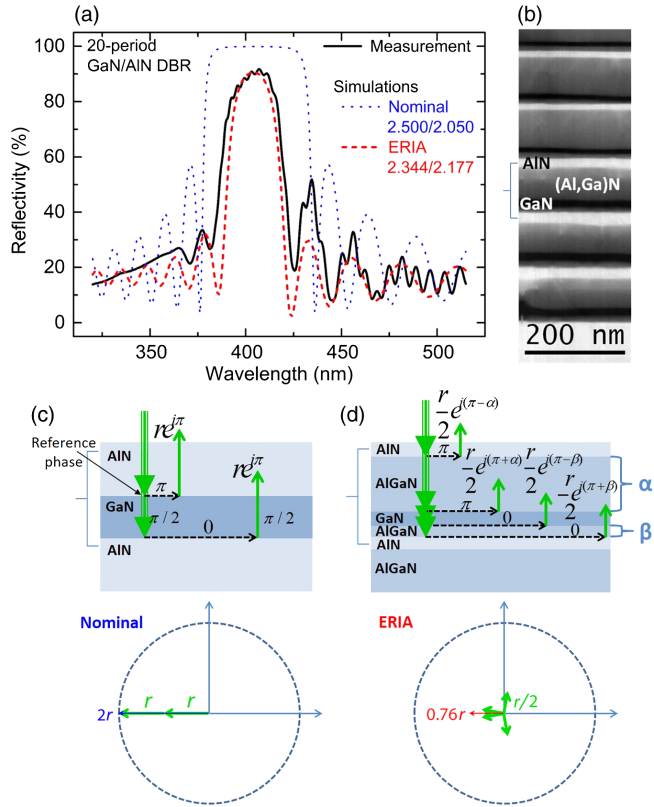

FIG. 3. (a) Reflectivity profile obtained from measurement and simulations: a nominal 20-period GaN/AlN DBR targeted at 405 nm and an equivalent standard-DBR counterpart determined by the ERIA method. (b) TEM image of the DBR cross section revealing strong intermixing at DBR interfaces. (c),(d) Sketches of the nominal and actual structures with their phasor diagrams according to the ERIA method [25].

TEM images of the resulting DBR reveal a highly periodic structure [Fig. 3(b)] [25]. However, strong intermixing at the DBR interfaces is observed, resulting in thick and nearly homogeneous (Al,Ga)N transient layers. The TL thicknesses are estimated at about 50 and 10 nm, respectively, with their phase thicknesses thus being approximately 100° and 20° [Fig. 3(b)]. The periodic formation of TL "weakens" the resonant condition, leading to the creation of reflected components which are no longer perfectly synchronized. Consequently, DBR optical properties deteriorate, as previously explained and experimentally observed in Fig. 3(a). The reflectivity profile of the resulting structure has a common "DBR-like" shape, but, however, significantly reduced stop-band parameters (height and width) with respect to the expected ones [see the nominal and measured reflectivity in Fig. 3(a)].

Making use of the phasor diagrams of the nominal and resulting structures [Fig. 3(c)] and previously derived formulas, the ideality factor of the structure is estimated at $I_F \approx 0.38$ ($D_F \approx 0.45$). Making use of the ERIA method, the resulting DBR with the fundamental parameters $n_{12} = 1.219$, $N = 20$, and $D_F = 0.45$ can be substituted for its standard-DBR counterpart (DBR') with the parameters $n'_{12} = 1.077$, $N' = 20$, and $D'_F = 0$.

Figure 3(a) shows the simulated reflectivity profile of the equivalent standard-DBR counterpart, with the parameters as calculated using the ERIA method. As can be seen, the result obtained via ERIA modeling fits very well the experimental reflectivity measurement, particularly concerning the stop-band height and width of the measured DBR and its simulated standard-DBR counterpart. It is worth noticing that the differences in the "short-wavelength" lateral fringes of the two structures arise from the absorption onset of the GaN layers whereas the differences in the "long-wavelength" lateral fringes are due to the "second-order Fabry-Perot reflectivity modulation" which arises at the GaN/sapphire interface. Neither of the two effects is considered in theoretical simulations, for simplicity.

## IV. ERIA AS A LINK BETWEEN STRUCTURAL DISORDER AND Q-FACTOR DETERIORATION OF RESONANT μCs

Before the ERIA method is applied to imperfect μCs, a brief summary of the μC Q factor is given, for clarity.

### A. Q factor of resonant μCs

The Q factor of an arbitrary resonator is defined as the total stored energy divided by the energy lost per 1 rad of the oscillating cycle. In the case of optical cavities, the total energy is directly proportional to the number of stored photons, with the Q factor thus being $2\pi$ times the inverse of the fraction of bouncing photons lost per full oscillating cycle. The origins of losses are leaks (L), occurring through the surrounding mirrors, and absorption (A), originating from the $\mu C$ layers. Consequently, the expected $Q$ factor can be decomposed into two contributing terms $Q^{-1} = Q_L^{-1} + Q_A^{-1}$, which are related to their corresponding losses in the following way: $L = 2\pi Q_L^{-1}$ and $A = 2\pi Q_A^{-1}$ [21]. In the further analysis, only an empty $\mu C$ is addressed. The lack of active region (most commonly realized in the form of quantum wells) reduces the amount of absorbed photons to a virtually zero value ($A \approx 0$ and $Q_A \approx \infty$), yielding $Q^{-1} = Q_L^{-1}$.

Structural imperfections of the resonant $\mu Cs$, such as layer intermixing and thickness variations, weaken the cavity resonant condition, affecting the synchronization of photons (otherwise perfectly synchronized) bouncing back and forth through the cavity medium. This desynchronization enhances photon leaks through the surrounding mirrors, leading to the subsequent deterioration of the $Q$ value. These additionally leaked photons are a direct consequence of undesired structural disorder.

For practical applications, planar $\mu Cs$ with high $Q$ factors are usually required. To achieve that, resonant cavity is sandwiched between two highly reflective DBRs, their targeted reflectivity ($R$) normally being approximately in the 0.99 range. For a comprehensive insight into a link between $\mu C$ structural disorders and the consequent $Q$ factor, a detailed understanding of the correlation between the structural and the optical disorder of high-reflectivity DBRs is first required. In the following section, this issue is addressed in detail.

### B. A link between structural and optical disorder of high-reflectivity DBRs

The DBR reflectivity ($R$) quantifies the fraction of reflected power, whereas DBR transmittivity ($T$) quantifies the fraction of transmitted power (both vs total incident power and at the targeted wavelength $\lambda$). The sum of the two parameters is equal to unity, $R + T = 1$, as long as DBR absorption is neglected (as in the present case). Switching to the corpuscular nature of light, the two parameters directly determine the fraction of reflected to transmitted photons on a DBR structure (at the targeted wavelength $\lambda$). Bearing, further, in mind that, when dealing with a high-reflectivity DBR ($R \approx 1$, i.e., $T \ll 1$), the following holds: $(n_0/n_s)[1/(n_{12}^{2N})] \ll 1$, allowing for further simplifications of reflectivity formula, given in Eq. (2) [the binomial approximation $(1 + x)^a \approx 1 + ax$ is used, where $|x| \ll 1$ and $a$ is an arbitrary real number]:

$$R = \left( \frac{1 - \frac{n_0}{n_s} \frac{1}{n_{12}^{2N}}}{1 + \frac{n_0}{n_s} \frac{1}{n_{12}^{2N}}} \right)^2 \approx \left( 1 - 4 \frac{n_0}{n_s} \frac{1}{n_{12}^{2N}} \right), \quad (13)$$

i.e.,

$$T \approx 4 \frac{n_0}{n_s} \frac{1}{n_{12}^{2N}}. \quad (14)$$

Structural imperfections of the resulting DBR weaken the resonant condition allowing additional (undesired) photon leaks through the structure. Substituting the resulting DBR ($n_{12}$, $N$, $D_F$) for its standard-DBR counterpart DBR' ($n'_{12}$, $N' = N$, $D'_F = 0$), the undesired leaks $D$ ($D$ stands for disorder) can be easily quantified:

$$D = T' - T \approx 4 \frac{n_0}{n_s} \left( \frac{1}{n'^{2N}_{12}} - \frac{1}{n_{12}^{2N}} \right)$$
$$= 4 \frac{n_0}{n_s} \frac{1}{n_{12}^{2N}} \left[ \left( \frac{1 + D_F n_{12}}{1 + \frac{D_F}{n_{12}}} \right)^{2N} - 1 \right], \quad (15)$$

where $T$ stands for transmission through the targeted DBR and $T'$ stands for transmission through the resulting (imperfect) DBR. Finally, the fraction of undesired to desired photon leak through a highly reflective DBR is given on the left-hand side of the equation,

$$\frac{D}{T} = \left[ \left( \frac{1 + D_F n_{12}}{1 + \frac{D_F}{n_{12}}} \right)^{2N} - 1 \right] \approx 2ND_F \left( n_{12} - \frac{1}{n_{12}} - D_F \right), \quad (16)$$

whereas its further simplified form (making use of the binomial approximation again) is given on the right-hand side. This simplified form holds for high-reflectivity DBRs ($T \ll 1$) with, in addition, relatively low structural disorder ($D_F \ll 1$). Note that Eq. (16) relates the optical disorder of the entire optical resonator—in this case, high-reflectivity DBR—to a structural disorder of its fundamental building block, i.e., its unit cell. It shows that disorder has a cumulative impact on the undesired to desired photon leak ratio ($D/T$). In the case of structures with a relatively low unit-cell disorder, the $D/T$ ratio increases nearly linearly with an increasing number of periods.

### C. A link between resonant $\mu C$ structural disorder and its $Q$ factor

In a standard planar $\mu C$, photons are stored in its middle part (in the resonant cavity itself) and leak through both the top and bottom mirrors (the following analysis addresses photon losses at the targeted wavelength $\lambda$). As previously stated, the $Q$ factor of an optical oscillator is $2\pi$ times the inverse of the fraction of bouncing photons lost per oscillating cycle: $L = 2\pi Q^{-1}$ (we remind the reader that photon losses due to absorption are neglected in the present case). In the specific case of planar $\mu Cs$, the fraction of photons lost per oscillating cycle can also be expressed as the ratio between the fraction of photons leaked through the surrounding mirrors over one "full cavity cycle," $(T + D)_{\text{top}} + (T + D)_{\text{bottom}}$, and the number of performed field oscillations during this cycle: $q$ (note that the photons bounce back and forth through the guiding medium, with one full cavity cycle consisting of a photon entire back-and-forth passage through the guiding medium). Combining the previous equations, we arrive at the equation which expresses the $Q$ factor as a function of the photon leaks (both desired and undesired) and the resonant cavity length $q$:

$$2\pi Q^{-1} = L = \frac{(T+D)_{\text{bottom}} + (T+D)_{\text{top}}}{q}. \quad (17)$$

We assume that the resonant cavity has a nominal optical thickness of $q_0\lambda$ (with $q_0$ being an integer), denoting it simply as $q_0$. Concerning the surrounding mirrors, we assume that they are characterized with the following parameters: $n_1$, $n_2$, $N$, and $D_F$, where $n_1$, $n_2$, and $N$ denote the DBR nominal parameters, i.e., the refractive indices ($n_1 > n_2$) and the number of periods, respectively, and $D_F$ denotes the disorder of the resulting DBR unit cells upon the $\mu C$ fabrication.

In planar $\mu$Cs, the photons bounce back and forth through the guiding medium. The back-and-forth bouncing is due to photon reflection at some of the surrounding DBRs' interfaces. The photons penetrate, to some extent, into surrounding DBRs, thus increasing the resonant-cavity optical thickness from its nominal value, $q_0$, to its effective $q$ value.

In the nominal (perfect) $\mu C$, the effective resonant-cavity thickness is estimated as [27]

$$q \approx q_0 + \left(\frac{n_{12}}{(n_{12}+1)(n_{12}-1)}\right)_{\text{bottom}} + \left(\frac{n_{12}}{(n_{12}+1)(n_{12}-1)}\right)_{\text{top}}. \quad (18)$$

Similarly, in the nominal structure, there are only desired photon losses $T(D=0)$, which amount to

$$T_{\text{bottom}} + T_{\text{top}} \approx \left(4\frac{n_0}{n_s}\frac{1}{n_{12}^{2N}}\right)_{\text{bottom}} + \left(4\frac{n_0}{n_s}\frac{1}{n_{12}^{2N}}\right)_{\text{top}}. \quad (19)$$

The combination of Eqs. (18) and (19) allows us to estimate the nominal $\mu C$ $Q$ factor:

$$2\pi Q^{-1} \approx \frac{\left(4\frac{n_0}{n_s}\frac{1}{n_{12}^{2N}}\right)_{\text{bottom}} + \left(4\frac{n_0}{n_s}\frac{1}{n_{12}^{2N}}\right)_{\text{top}}}{q_0 + \left(\frac{n_{12}}{(n_{12}+1)(n_{12}-1)}\right)_{\text{bottom}} + \left(\frac{n_{12}}{(n_{12}+1)(n_{12}-1)}\right)_{\text{top}}}. \quad (20)$$

In the resulting (imperfect) $\mu C$, DBR unit cells have nonzero disorder ($D_F > 0$), i.e., the DBR effective refractive-index ratio is reduced with respect to the nominal one: $n'_{12} < n_{12}$. The disorder weakens the resonant condition, allowing (undesired) photon leaks through the surrounding mirrors. It also allows a somewhat deeper penetration of the bouncing photons into the surrounding mirrors. It is worth noticing that the influence of disorder on these two terms is very different. While the former increase (undesired photon leaks) can be very significant, the latter increase (effective cavity length) is normally only very slight; the combination of the two may lead to, in total, a substantial $Q$-factor decrease.

To estimate the actual $Q$ factor ($Q'$) of the resulting (imperfect) $\mu C$, the ERIA method can be applied to the surrounding mirrors. For that, the nominal refractive-index ratios $n_{12}$ of the surrounding mirrors are substituted with the effective ones $n'_{12}$:

$$2\pi Q'^{-1} \approx \frac{\left(4\frac{n_0}{n_s}\frac{1}{n'^{2N}_{12}}\right)_{\text{bottom}} + \left(4\frac{n_0}{n_s}\frac{1}{n'^{2N}_{12}}\right)_{\text{top}}}{q_0 + \left(\frac{n'_{12}}{(n'_{12}+1)(n'_{12}-1)}\right)_{\text{bottom}} + \left(\frac{n'_{12}}{(n'_{12}+1)(n'_{12}-1)}\right)_{\text{top}}}. \quad (21)$$

Similarly, making use of Eq. (7), the $Q$ factor of the resulting $\mu C$ is expressed only as a function of the cavity length ($q_0$) and the fundamental parameters of the surrounding DBRs ($n_{12}$, $N$, $D_F$):

$$2\pi Q'^{-1} \approx \frac{\left[4\frac{n_0}{n_s}\left(\frac{1}{n_{12}}\frac{1+D_F n_{12}}{1+\frac{D_E}{n_{12}}}\right)^{2N}\right]_{\text{bottom}} + \left[4\frac{n_0}{n_s}\left(\frac{1}{n_{12}}\frac{1+D_F n_{12}}{1+\frac{D_E}{n_{12}}}\right)^{2N}\right]_{\text{top}}}{q_0 + \left(\frac{(n_{12}+D_F)(1+D_F n_{12})}{(n_{12}+1)(n_{12}-1)(1-D_F^2)}\right)_{\text{bottom}} + \left(\frac{(n_{12}+D_F)(1+D_F n_{12})}{(n_{12}+1)(n_{12}-1)(1-D_F^2)}\right)_{\text{top}}}. \quad (22)$$

Finally, it is worth noticing that, when dealing with $\mu$Cs with a relatively low disorder ($D_F \ll 1$), the binomial approximation can be further applied:

$$2\pi Q'^{-1} \approx \frac{4\frac{n_0}{n_s}\frac{1}{n_{12}^{2N}}\left[1+2ND_F\left(n_{12}-\frac{1}{n_{12}}\right)\right]_{\text{bottom}} + 4\frac{n_0}{n_s}\frac{1}{n_{12}^{2N}}\left[1+2ND_F\left(n_{12}-\frac{1}{n_{12}}\right)\right]_{\text{top}}}{q_0 + \left(\frac{n_{12}+D_F(1+n_{12}^2)}{(n_{12}+1)(n_{12}-1)}\right)_{\text{bottom}} + \left(\frac{n_{12}+D_F(1+n_{12}^2)}{(n_{12}+1)(n_{12}-1)}\right)_{\text{top}}}. \quad (23)$$

A comparison of Eqs. (20) and (23) clearly reveals (and quantifies) the influence of structural unit-cell disorder on final $Q$-factor deterioration. More precisely, it predicts that, despite low disorder ($D_F \ll 1$), the increase in photon leaks can be significant [see the numerator in Eq. (23)] since the impact of unit-cell disorder on it is cumulative; the disorder is multiplied by the number of DBR periods, which is typically large for high-$Q$-factor planar $\mu C$s. On the other hand, it predicts only a very slight increase in the effective cavity length [see the denominator in Eq. (23)] since the impact of unit-cell disorder on it is not cumulative. In this case, the disorder is multiplied by a constant $(1 + n_{12}^2)$, which is typically in the 2–4 range.

### D. Theoretical validation of formulas relating $\mu C$ disorder and its $Q$ factor

For the theoretical validation of the ERIA method applied on imperfect $\mu C$s, a $\mu C$ consisting of a $3\lambda$ resonant GaN cavity sandwiched between two GaN/AlN DBRs, centered at $\lambda = 405$ nm and fabricated on a GaN template, is chosen. The bottom (top) DBRs both have 20 (20) periods, and their nominal refractive-index ratio is 2.50/2.05. We suppose that the GaN/AlN interfaces are smeared, creating linearly graded transient layers. The phase thicknesses of all transient layers are set at $\alpha = \beta = 60° = \pi/3$. The refractive-index profiles of the targeted and the resulting structure are shown in Fig. 4(a) (only the resonant cavity with the three surrounding DBR periods are shown, for clarity).

Figure 4(b) compares the reflectivity profile of the "real" structure [as shown in Fig. 4(a)] to that obtained on the equivalent $\mu C$, the parameters of which have been determined via the ERIA method. The equivalent $\mu C$ consists of a $3\lambda$ GaN resonant cavity, sandwiched between two standard DBRs with effective refractive-index ratios of 2.457/2.085 (see the linearly graded transient layers in Table I, for clarity). The reflectivity profiles of the real structure and that obtained for the ERIA-equivalent structure are in virtually perfect agreement. The only relevant difference concerns a slight shift of the resonant dip position. In the present case, the resonant wavelength is set at 405 nm, and the resonant dip redshifts 0.78 nm (less than 0.2%), when the ERIA method is applied [28]. The dip shape, which is tightly related to the resonant cavity $Q$ factor, is closely preserved [see Fig. 4(c)].

To estimate the accuracy of the ERIA method in determining the $Q$ factor of disordered $\mu C$s, we compare results obtained for the real $\mu C$ with those obtained on its equivalent $\mu C$, as determined by the ERIA method. Table II summarizes $Q$-factor values (for the previously described $\mu C$ with linearly graded transient layers) when the number of bottom (top) DBR periods vary from 10 (10) to 40 (40). Column $A$ makes direct use of Eq. (22) applied on the ERIA-equivalent $\mu C$, the parameters of which can be found in Table I (both the bottom and top ERIA-equivalent DBRs have $n_1 = 2.457$, $n_2 = 2.085$, $n_{12} = 1.178$, $I_F = 0.827$, $D_F = 0.095$, and $N = 10, 15, …, 40$). The $Q$ factor of the cavity is numerically determined using two additional methods. In the first method, we consider the angular frequency $\omega$ of the propagating wave to be a complex variable $\omega = \omega_r + i\omega_i$, and we find it to be a root of the (2,2) element of the transfer matrix. The $Q$ factor of both real $\mu C$ and its ERIA-equivalent $\mu C$ is then calculated as $Q = [(\omega_r)/(2\omega_i)]$ (see Appendix B for details) and represented in columns B and C of Table II. In the second method, we calculate the dependence of the reflectivity on the wavelength using the TMS and estimate the $Q$ factor from full width at half maximum $\Delta\lambda$ of the reflectivity dip [see, for example, Fig. 4(b)] at the central wavelength $\lambda$ as $Q = [\lambda/(\Delta\lambda)]$. The $Q$ factors of both the real $\mu C$ and its ERIA-equivalent $\mu C$ are represented in columns D and E of Table II. The results from Table II are also graphically featured in Fig. 4(d). They show that the ERIA method can be very successfully applied for the $Q$-factor determination of the disordered $\mu C$s. In particular, we emphasize that the $Q$ factor can be directly assessed via Eq. (22).

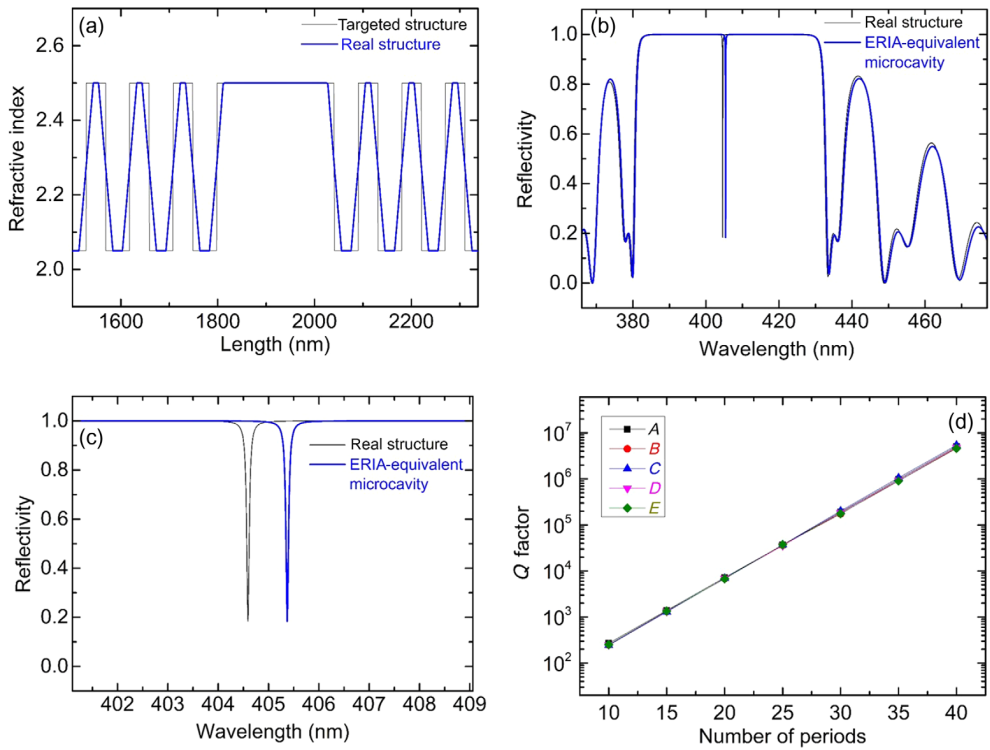

FIG. 4. (a) Refractive-index profiles of a $3\lambda$ resonant cavity sandwiched between two 20-period GaN/AlN DBRs. Both targeted and resulting (real) structures (obtained when linearly graded transient layers are formed at the DBR heterointerfaces) are shown. (b) Reflectivity of the real $\mu C$ and its ERIA equivalent $\mu C$, both assessed via TMSs. (c) The two reflectivity dips in detail. (d) Comparison of $Q$ factors estimated by the methods specified in Table II.

TABLE II. $Q$ factor estimated on a real structure and its ERIA-equivalent $\mu C$. Column A: Making use of Eq. (22) on the ERIA-equivalent $\mu C$. Columns B and C: Making use of the $\omega_r/2\omega_i$ value on real and ERIA-equivalent structures, respectively. Columns D and E: Making use of the $\lambda/\Delta\lambda$ value on real and ERIA-equivalent structures, respectively.

| Number of DBRs' periods | | Q factor | | | | |
|---|---|---|---|---|---|---|
| | | A | B | C | D | E |
| Bottom | Top | Equation (22) on equivalent ERIA cavity | $\omega_r/2\omega_i$ on real cavity | $\omega_r/2\omega_i$ on equivalent ERIA cavity | $\lambda/\Delta\lambda$ on real cavity | $\lambda/\Delta\lambda$ on equivalent ERIA cavity |
| 10 | 10 | 270 | 248 | 244 | 252 | 253 |
| 15 | 15 | 1392 | 1320 | 1300 | 1330 | 1350 |
| 20 | 20 | 7174 | 6840 | 6905 | 6860 | 6750 |
| 25 | 25 | 36 963 | 36 700 | 37 300 | 35 800 | 37 500 |
| 30 | 30 | 190 451 | 174 000 | 202 000 | 184 000 | 176 000 |
| 35 | 35 | 981 288 | 931 000 | 1 060 000 | 933 000 | 900 000 |
| 40 | 40 | 5 056 030 | 4 720 000 | 5 390 000 | 4 820 000 | 4 600 000 |

## V. DISCUSSION

### A. Influence of disorder on peak reflectivity, stop-band width and $Q$ factor

In the previous sections, the most relevant parameters of the planar optical resonators are analytically linked to unit-cell disorder. These results are further graphically presented in Fig. 5.

Figure 5(a) shows the decreases in peak reflectivity and stop-band width of a 20-period GaN/AlN DBR as a function of increasing unit-cell disorder. Note that, in the case of high-reflectivity DBRs, the disorder more strongly affects the stop-band width than the DBR peak reflectivity itself. The background reason for this trend is that, in the ideal structure, the transmission of photons through the DBR is exceptionally low (in the previous case it is estimated at $T \approx 6 \times 10^{-4}$) and, despite a dramatic (relative) increase in undesired photon leaks through the DBR structure, the portion of leaked photons becomes comparable to the portion of reflected photons only at relatively high disorder values. Note that the undesired photon leaks increase as a polynomial function, with its degree being $2N$, i.e., 40 in the present case [see Eq. (15)]. The stop-band width, on the other hand, is directly proportional to the unit-cell ideality factor, with the stop-band deterioration thus being nearly linear with increasing unit-cell disorder [see Eq. (10)].

Figure 5(b) summarizes the $Q$-factor deterioration of a $3\lambda$ resonant GaN cavity, sandwiched between two 20-period GaN/AlN DBRs, as a function of increasing disorder making use of Eqs. (22) and (23). Note first that the binomial approximation used in Eq. (23) yields a good $Q$-factor estimation for low unit-cell disorder ($D_F < 0.1$), as expected. Since the $Q$ factor is directly linked to the portion of undesirably leaked photons, its deterioration with increasing disorder is very fast. For quantitative comparison, at $D_F = 0.1$, the corresponding DBR peak reflectivity and stop-band width [Fig. 5(a)] decrease approximately 0.2% and about 10%, respectively, whereas the corresponding $Q$ factor [Fig. 5(b)] decreases approximately 73%, respectively.

As commented, the $Q$-factor deterioration is tightly linked to the portion of undesirably leaked photons, and this portion is further directly linked to the number of DBRs periods used to "sandwich" the resonant cavity. To demonstrate this trend, $Q$ factors of a $3\lambda$ resonant GaN cavity sandwiched between ideal ($D_F = 0$) and disordered ($D_F = 0.1$) GaN/AlN DBRs with a varying number of periods are also depicted in Fig. 5(c) [the $Q$ values are estimated via Eq. (22)]. This graph confirms that the same disorder ($D_F = 0.1$) has a much stronger impact on deterioration of a resonant cavity with a higher number of periods. The quality factor of a disordered ($Q'$) vs a targeted ($Q$) cavity ($D_F = 0.1$ vs 0) falls to $Q' \approx 0.55Q$ and $Q' \approx 0.06Q$ values, for $N = 10$ and $N = 40$, respectively. The much higher deterioration of the $Q$ factor in the latter case is directly linked to the cumulative influence that disorder has on the resonant-cavity $Q$ factor, as previously commented.

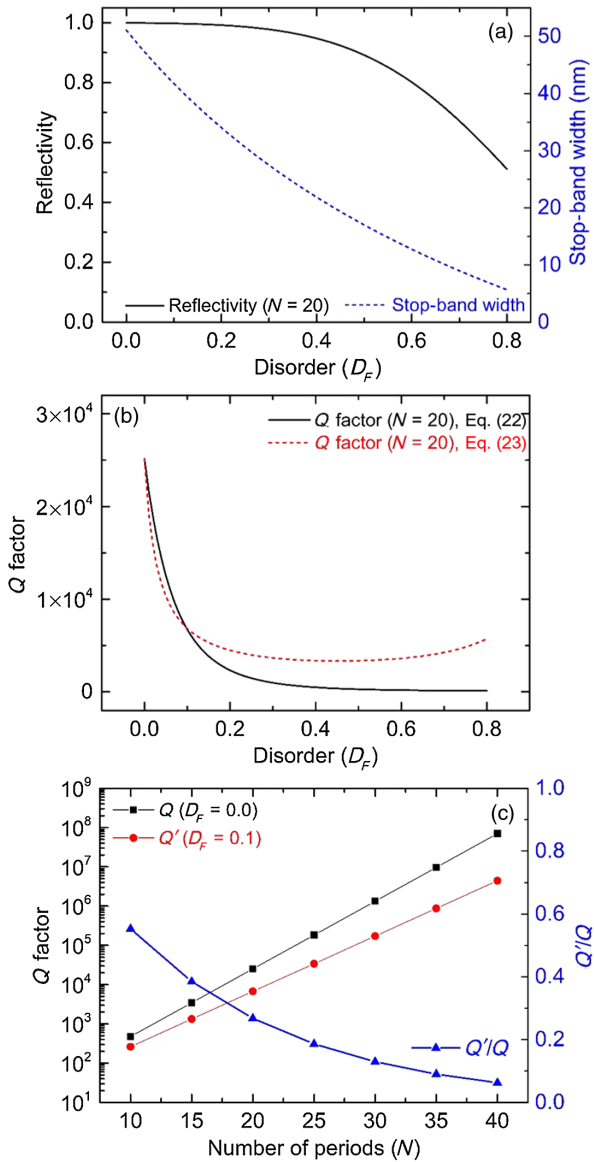

FIG. 5. Influence of disorder ($D_F$) on optical performance of GaN/AlN planar optical resonant structures. (a) Decrease in peak reflectivity [Eq. (9)] and stop-band width [Eq. (10)] of a 20-period GaN/AlN DBR as a function of increasing disorder. (b) Decrease in $Q$ factor of a $3\lambda$ resonant GaN cavity sandwiched between two 20-period GaN/AlN DBRs, as a function of increasing disorder [Eqs. (22) and (23)]. (c) $Q$ factors of a $3\lambda$ resonant GaN cavity sandwiched between ideal ($D_F = 0$) and disordered ($D_F = 0.1$) GaN/AlN DBRs with a varying number of periods [Eq. (22)].

Finally, let us note that the results presented in Fig. 5(a) very likely explain experimental reflectivity measurements obtained for a wide variety of disordered DBRs reported in the literature; similarly, results reported in Figs. 5(b) and 5(c) explain the experimental $Q$-factor measurements (in particular, poor agreement between the experiment and theory and strong variations in $Q$ factor across the cavity surface, obtained in microreflectivity experiments) reported for a wide variety of resonant μCs [18,19,21,25,29–35].

### B. Limitations of the ERIA method

The maximization of a DBR reflectivity relies on design which provides perfect synchronization of reflected components. The undesired DBR structural disorder leads to desynchronization of the reflected components, consequently leading to a decrease in reflectivity. As previously explained, the ERIA method is based on a simple observation that the same effect, i.e., the decrease in reflectivity, is obtained when the DBR refractive-index ratio is properly reduced. As a consequence, the optical performance of a disordered DBR is easily quantified, simply by reducing its refractive-index ratio to its effective value ($n_{12} \rightarrow n'_{12}$) while keeping its structural ideality.

The ERIA method departs from the precise quantitative definition of the structural ideality ($I_F$), i.e., disorder ($D_F$), of a single DBR unit cell. Then, assuming that the disorder is preserved throughout the reflector, the method quantifies the DBR optical deterioration. The strictly periodic refractive-index profile, assumed at the beginning, however, is a strong condition that does not have to be satisfied for successful application of the ERIA method. In that respect, the applicability of the method is further discussed below.

Note that the preservation of the DBR unit-cell disorder [see Eqs. (4) and (5)] explicitly assumes that the waves reflected on the resulting (imperfect) DBR unit cells have (i) the same phase—with the phases matching the phase of the wave reflected on the targeted (perfect) unit cell—and (ii) the same intensity—with the intensities being lower than that of the wave reflected on the targeted (perfect) unit cell.

The fulfillment of conditions (i) and (ii) is a necessary and sufficient prerequisite to guarantee preservation of unit-cell disorder $D_F$ throughout the DBR structure, and thus a proper implementation of the ERIA method. As can be seen, the condition of the periodically changing refractive-index profile assumed above is stricter than the actual required prerequisite.

Concerning the fulfillment of condition (i), it is easy to see that it requires the conservation of the DBR unit-cell optical thickness throughout the entire structure. Note that the optical thickness of quarter-wave layers is directly controlled via their thicknesses. The variations in unit-cell optical thicknesses throughout a DBR are not common in standard technological processes (except for rough growth errors, such as mistaken growth times, etc.) [25]. Anyway, it is worth noticing that varying optical thicknesses of DBR periods implies varying phases of reflected waves (from one DBR period to another); bearing the phasor diagram of the structure in mind (Fig. 1), it is clear that this structural disorder type leads to varying disorder from one period to another ($D_F \neq $ const). Consequently, this type of structure is not convenient for direct application of the ERIA method, as exposed in this article; further comments concerning some cases of DBRs with $D_F \neq \text{const}$ and applicability of the ERIA method are included below. The common (intrinsic) origins of the unit-cell structural disorder are either rough interfaces (a consequence of random thickness fluctuations) or smeared interfaces (a consequence of the two materials intermixing) [6]. On average, neither of the two deterioration mechanisms changes the unit-cell optical thickness, implying that condition (i) is commonly fulfilled in real structures.

The fulfillment of (ii) in real optical resonators is, on the other hand, more critical and less clear than the fulfillment of (i). Semiconductor DBRs are normally constituted of two materials with different lattice constants. The lattice mismatch leads to a strain accumulation in an early growth stage, with the strain then being progressively relaxed during the growth via defect formation [36,37]. The DBR layers are thus under alternatively tensile and compressive strain (i.e., positive and negative stress). The strain relaxation leads to an overall stress reduction in the structure, with the average stress eventually reaching zero for sufficiently thick DBRs; from this point on, the local stress continues to oscillate around zero, while preserving a nearly zero average value. Thus, unlike (i), the fulfillment of (ii) strongly depends on the properties of the two materials applied for DBR fabrication and can vary significantly from one material system to another.

We refer to the part of the epitaxial growth during which the stress in the structure is being reduced as a growth transient. During this initial growth stage, significant structural changes may happen in the DBR since strain relaxation mechanisms (i.e., defect formation) are active. The defect formation leads to structural damage and, most likely, to a progressive increase in unit-cell disorder ($D_F\uparrow$). Once this transient is finished, it is very likely that the structural disorder will remain nearly constant, reaching a certain (asymptotic) value. Note that this directly implies that, unlike its bottom part, the DBR upper part will have a nearly constant disorder and will thus be appropriate for ERIA modeling.

Practical applications normally require high-reflectivity DBRs, i.e., DBRs with a high number of periods. This implies that a very small fraction of incident photons reach the DBR bottom part (which is inconvenient for the ERIA method) but reflect within its top part (which is convenient for the ERIA method). Based thus on the previous reasoning, we speculate that, despite the varying disorder ($D_F$) of the real DBRs, the applicability of the ERIA method may still be very high. In the particular case of disordered GaN/AlN DBRs that we fabricate (with a total of 15 DBRs; results are not presented), which contain between 6 and 30 periods and have targeted wavelengths varying in the (380–550)-nm range, the ERIA method proves to be an exceptionally good analysis tool [25]. Note that once the reflectivity measurement is performed, the $I_F$ and $D_F$ factors are directly estimated from the measured (vs nominal) stop-band width [see Eq. (10), for clarity]. If the substitution of the $D_F$ (determined from the reduction of the stop-band width) in the equation for peak reflectivity [Eq. (9)] yields a peak reflectivity that is in good agreement with the measured one, then the disorder of the entire DBR can be considered nearly constant and, consequently, the (disordered) DBR under study can be substituted with its ERIA-equivalent counterpart [the fingerprint of a DBR with a nearly constant disorder is (i) the reflectivity profile shape matches the reflectivity profile shape of a standard DBR, and (ii) the stop-band height and width are both reduced with respect to the expected nominal value]. We nevertheless point out that the applicability of the method could vary significantly from one material system to another and that additional quantitative analyses and their comparison with experimental results are necessary to address this issue with clarity.

## VI. CONCLUSION

In summary, this work provides a detailed insight into a link between the structural and the optical disorder of planar resonant optical structures, in particular, DBRs and resonant $\mu$Cs. The link between the two is unraveled by making use of the ERIA method, which shows that the optical deterioration of a DBR (originating from its structural imperfections) can be modeled simply by reducing the unit-cell refractive-index ratio ($n_{12} \rightarrow n'_{12}$) while keeping its structural ideality. We first propose a precise quantitative definition of the DBR unit-cell ideality and disorder. Then, making use of the ERIA method, we show that the optical response of any DBR with a periodic refractive-index profile can be very well approximated using its corresponding standard-DBR counterpart. We further derive the equivalent standard-DBR structures for the cases involving DBRs with homogeneous, linearly graded, sine-wave graded, biparabolically graded, uniparabolically graded, and interdiffused TLs at interfaces. The results are validated making use of both TMSs and direct experimental measurements of imperfect DBRs. In the second part of this article, the ERIA method is further applied on resonant $\mu$Cs, unraveling the link between their structural disorder and the subsequent deterioration of their $Q$ factors. The obtained results are validated via TMSs. The analytical formulas derived in this article enable rapid insight (both quantitative and qualitative) into optical properties of imperfect DBRs and $\mu$Cs. This work establishes a base to understand the link between structural disorder and the subsequent deterioration in performance of planar optical resonators.

## ACKNOWLEDGMENTS

This work was supported by a Short Term Scientific Mission Grant from COST Action MP1406. N. V. gratefully acknowledges the support from the Ministry of Education, Science and Technological Development of the Republic of Serbia (Project No. ON171017) and the European Commission under H2020 project VI-SEEM, Grant No. 675121.

## APPENDIX A: IDEALITY FACTOR DETERMINATION

In this appendix, making use of Eq. (12), ideality factors for DBR unit cells presented in Table I are derived. The corresponding refractive-index profiles $n(\varphi)$ are depicted and analytically expressed in Table I, for clarity. Their first derivatives $n'(\varphi)$ are mathematically expressed in the equations below as parts of the corresponding integrals. In the following equations, $\delta(\varphi)$ stands for the Dirac $\delta$ function.

Standard DBR unit cell (no TLs):

$$I_F = \frac{1}{2(n_1-n_2)} \int_{0_-}^{\pi_-} \left[(n_1-n_2)\delta(\varphi) - (n_1-n_2)\delta\left(\varphi - \frac{\pi}{2}\right)\right] e^{2i\varphi} d\varphi = \frac{1}{2}(1+1) = 1. \tag{A1}$$

Homogeneous TLs:

$$I_F(\alpha,\beta) = \frac{1}{2(n_1-n_2)} \int_{-(\alpha/2_-)}^{\pi-(\alpha/2_-)} \left\{\left(\frac{n_1-n_2}{2}\right)\left[\delta\left(\varphi+\frac{\alpha}{2}\right)+\delta\left(\varphi-\frac{\alpha}{2}\right)\right] - \left(\frac{n_1-n_2}{2}\right)\left[\delta\left(\varphi-\frac{\pi}{2}-\frac{\beta}{2}\right)+\delta\left(\varphi-\frac{\pi}{2}+\frac{\beta}{2}\right)\right]\right\} e^{2i\varphi} d\varphi$$
$$= \frac{1}{2}(\cos\alpha + \cos\beta). \tag{A2}$$

Linearly graded TLs:

$$I_F(\alpha,\beta) = \frac{1}{2(n_1-n_2)} \left(\int_{-(\alpha/2)}^{\alpha/2} \frac{n_1-n_2}{\alpha} e^{2i\varphi} d\varphi + \int_{(\pi/2)-(\beta/2)}^{(\pi/2)+(\beta/2)} -\frac{n_1-n_2}{\beta} e^{2i\varphi} d\varphi\right) = \frac{1}{2}\left(\frac{\sin\alpha}{\alpha} + \frac{\sin\beta}{\beta}\right). \tag{A3}$$

Sine-wave graded TLs:

$$I_F(\alpha,\beta) = \frac{1}{2(n_1-n_2)} \left\{\int_{-(\alpha/2)}^{\alpha/2} \frac{\pi}{\alpha} \frac{n_1-n_2}{2} \cos\left(\frac{\pi}{\alpha}\varphi\right) e^{2i\varphi} d\varphi + \int_{(\pi/2)-(\beta/2)}^{(\pi/2)+(\beta/2)} -\frac{\pi}{\beta} \frac{n_1-n_2}{2} \cos\left[\frac{\pi}{\beta}\left(\varphi - \frac{\pi}{2}\right)\right] e^{2i\varphi} d\varphi\right\}$$
$$= \frac{1}{2}\left(\frac{\cos\alpha}{1-4\alpha^2/\pi^2} + \frac{\cos\beta}{1-4\beta^2/\pi^2}\right). \tag{A4}$$

### 1. Biparabolically graded TLs

It is worth noticing that, in all previous cases, two conditions hold: First, the reference phase $\varphi = 0$ of the $n(\varphi)$ function is set in the middle of the first TL; second, the refractive-index first derivative $n'(\varphi)$ is an even function [see Eqs. (A1)–(A4)]. These two conditions provide that the imaginary parts of Eqs. (A1)–(A4) cancel out, yielding positive real numbers as equation solutions, i.e., ideality factors. In the case of parabolically graded TLs, the $n'(\varphi)$ is not an even function and a proper setting of the reference phase (which would yield canceling of the imaginary terms in the corresponding equation), under this condition, is not trivial. Nevertheless, it is easy to show that the ideality factor is always (i.e., for an arbitrarily set reference phase) equal to the modulus of the complex number obtained as a result via Eq. (12). The integration of the first parabolically graded TL, according to its analytical form featured in Table I and Eq. (12) yields

$$\frac{1}{2(n_1-n_2)} \int_0^\alpha 2\frac{n_1-n_2}{\alpha^2}(\alpha-\varphi) e^{2i\varphi} d\varphi$$
$$= \frac{1}{4\alpha^2}(-e^{2i\alpha} + 2i\alpha + 1) \tag{A5}$$

and

$$\left|\frac{1}{4\alpha^2}(-e^{2i\alpha} + 2i\alpha + 1)\right| = \frac{1}{2}\frac{\sqrt{\alpha^2 - \alpha\sin(2\alpha) + (\sin\alpha)^2}}{\alpha^2}. \tag{A6}$$

Supposing that the resulting waves reflected at the first and second interfaces are perfectly synchronized (i.e., in phase), we find the ideality factor of the biparabolically graded unit cell:

$$I_F(\alpha,\beta) = \frac{1}{2}\left(\frac{\sqrt{\alpha^2 - \alpha\sin(2\alpha) + (\sin\alpha)^2}}{\alpha^2} + \frac{\sqrt{\beta^2 - \beta\sin(2\beta) + (\sin\beta)^2}}{\beta^2}\right). \quad (A7)$$

### 2. Uniparabolically graded TLs

This type of unit cell contains one parabolically graded TL and one sharp interface, with no TLs. If the waves reflected at the two interfaces are perfectly synchronized, then

$$I_F(\alpha, \beta = 0) = \frac{1}{2}\left(\frac{\sqrt{\alpha^2 - \alpha\sin(2\alpha) + (\sin\alpha)^2}}{\alpha^2} + 1\right). \quad (A8)$$

### 3. Interdiffused TLs

According to Ref. [26], the refractive-index profile of DBRs with interdiffused TLs has the following form:

$$n(\varphi) = \frac{n_1 + n_2}{2} + \frac{2}{\pi}(n_1 - n_2)\sum_{m=1}^{\infty}\frac{1}{m}\cos(m\pi)\sin\left(\frac{m\pi}{2}\right)$$
$$\times \cos(2m\varphi)e^{-m(4\pi\bar{n}L_d/\lambda)^2}. \quad (A9)$$

It is easy to show that only the first component in the previous sum ($m = 1$) yields a nonzero integral. After arranging the phase of this term to match the reference phase, as defined in the derivation of Eq. (12) [the new phase is shifted $+(\pi/4)$, i.e., $\varphi$ is substituted with $[\varphi + (\pi/4)]$], we get

$$I_F(L_d) = \frac{1}{2(n_1 - n_2)}\int_0^\pi \frac{4}{\pi}(n_1 - n_2)\cos(2\varphi)$$
$$\times e^{-(4\pi\bar{n}L_d/\lambda)^2}e^{2i\varphi}d\varphi$$
$$= e^{-(4\pi\bar{n}L_d/\lambda)^2}. \quad (A10)$$

## APPENDIX B: DETERMINATION OF THE MICROCAVITY $Q$ FACTOR FROM THE TRANSFER MATRIX OF THE STRUCTURE

In this appendix, the method for calculating the microcavity $Q$ factor from the transfer matrix of the structure is derived. We consider a planar microcavity grown along the $x$ axis. The electric field (more precisely its complex representative) in the region above the cavity (in air, for example) is generally given by the expression

$$E_a(x,t) = E_a^+ e^{i(\omega t - k_a x)} + E_a^- e^{i(\omega t + k_a x)}, \quad (B1)$$

while, in the region below the cavity (in the substrate, for example), it reads

$$E_b(x,t) = E_b^+ e^{i(\omega t - k_b x)} + E_b^- e^{i(\omega t + k_b x)}. \quad (B2)$$

In Eqs. (B1) and (B2), the $x$ axis is directed from the air toward the substrate, $E_a^+$ is the amplitude of the wave entering the cavity from the air, $E_a^-$ is the amplitude of the wave leaking from the cavity into the air, and $E_b^+$ is the amplitude of the wave leaking from the cavity into the substrate, while $E_b^-$ is the amplitude of the wave entering the cavity from the substrate, $\omega = \omega_r + i\omega_i$ is the angular frequency of the wave which may also contain an imaginary part that describes the time decay of the wave, while $k_a$ and $k_b$ are the wave vectors which may also contain an imaginary part.

The transfer matrix $M$ relates the amplitudes of electric fields of incoming and outgoing waves as

$$\begin{bmatrix} E_b^+ \\ E_b^- \end{bmatrix} = \begin{bmatrix} M_{11} & M_{12} \\ M_{21} & M_{22} \end{bmatrix} \begin{bmatrix} E_a^+ \\ E_a^- \end{bmatrix}. \quad (B3)$$

In the case of electromagnetic-field oscillations in the cavity, boundary conditions $E_a^+ = 0$ and $E_b^- = 0$ are satisfied. After the substitution of these boundary conditions in Eq. (B3), one obtains that oscillations of electromagnetic field can be sustained only if $M_{22} = 0$. From this condition, we find the complex angular frequency $\omega$. Its real part determines the wavelength of the oscillating field, while its imaginary part determines the time decay of the field. The definition of the $Q$ factor reads

$$Q = 2\pi \frac{W}{\Delta W}, \quad (B4)$$

where $W$ is the total energy of the field and $\Delta W$ is its loss during one period $T$ of oscillation. Since the energy of the field is proportional to the square of its amplitude, we obtain $W \sim e^{-2\omega_i t}$ and, consequently,

$$Q = 2\pi \frac{1}{1 - e^{-2\omega_i T}}. \quad (B5)$$

Since $\omega_i T \ll 1$ is satisfied in typical cavities, the last equation reduces to

$$Q = \frac{\omega_r}{2\omega_i}, \quad (B6)$$

which is the expression that we use to calculate the $Q$ factor.

---